\documentclass[aps,prl,twocolumn,superscriptaddress,showpacs,floatfix]{revtex4}
\usepackage{graphicx}
\usepackage{bm}
\begin{document}


\title{Ingredients for an efficient thermal diode}

\author{Shunda Chen}
\affiliation{Center for Nonlinear and Complex Systems,
Universit\`a degli Studi dell'Insubria, via Valleggio 11, 22100 Como, Italy}
\affiliation{Istituto Nazionale di Fisica Nucleare, Sezione di Milano,
via Celoria 16, 20133 Milano, Italy}

\author{Emmanuel Pereira}
 \email{emmanuel@fisica.ufmg.br}
\affiliation{Departamento de F\'{\i}sica--Instituto de Ci\^encias Exatas, Universidade Federal de Minas Gerais, CP 702 30.161-970 Belo Horizonte MG, Brazil}

\author{Giulio Casati}
 \email{Giulio.Casati@uninsubria.it}
\affiliation{Center for Nonlinear and Complex Systems,
Universit\`a degli Studi dell'Insubria, via Valleggio 11, 22100 Como, Italy}
\affiliation{Istituto Nazionale di Fisica Nucleare, Sezione di Milano,
via Celoria 16, 20133 Milano, Italy}
\affiliation{International Institute of Physics, Federal University of Rio Grande do Norte, Natal, Brazil}
\date{\today}

\begin{abstract}

We provide convincing empirical evidence that long range interactions strongly enhance the rectification effect which takes place in mass graded systems. Even more importantly the rectification does not decrease with the increase of the system size. Large rectification is obtained also for the equal mass case and with graded on-site potential. These results allow to overcome current limitations of the rectification mechanism and open the way for a realistic implementation of efficient thermal diodes.

\end{abstract}



\maketitle


The great development of modern electronics with the invention of electric diodes, transistors and other nonlinear solid-state devices has led to
impressive
scientific and technological achievements, with strong impact in our daily lives.
However its counterpart, aimed at the manipulation of the heat current and phononics, more than a decade after the first proposal of the thermal diode \cite{Casati1},
remains at a near-standstill in terms of management of thermal circuits, basically due to the lacking  of efficient and feasible diodes.

  The thermal diode is the basic tool for the
manipulation and control of the heat flow. In this device heat flows preferably in one direction, i.e., the magnitude of
the heat current changes as we invert the sample between two thermal baths. The most recurrent proposals of thermal diodes are based on the sequential coupling
of two or three segments with anharmonic interactions \cite{Casati1, Casati2, BHu1} and they have to face several problems which are difficult to overcome. In particular
their rectification power is small and rapidly decays to zero as the system size increases. For this reason, more and more efforts have been devoted to the search of alternative systems or rectifying mechanisms, however without remarkable results.

In this letter, we suggest a possible way to overcome this conundrum, namely we identify general and simple ingredients to build a diode with  great rectification effect which moreover does not decay to
zero with increasing the system size. By considering one dimensional anharmonic chains of oscillators, we provide empirical evidence that
graded mass distribution
and long range interparticle interactions, lead to a substantial improvement of the thermal rectification phenomenon.
It is interesting that a comparable  very large rectification factor is also obtained if asymmetry in the system is introduced in a different way, e.g., by
considering the equal mass case with graded on-site potential.

We would like to stress the following points. First, since the pioneering work of Debye, anharmonic chains of oscillators are the most basic and recurrent models for the study of heat conduction in insulating crystals as they are believed to mimic, to some extent, the behavior of real materials.

Second, we recall that graded materials, i.e., inhomogeneous systems whose structure changes gradually in space, are not only abundant in nature, but they can also
be manufactured. They have attracted considerable interest in many areas: material science, engineering, optics, etc. \cite{Huang}. In particular, a simple type of graded diode has been already experimentally built \cite{Chang}: it consists of a carbon and boron nitride nanotube, inhomogeneously mass-loaded with heavy molecules. Unfortunately, the rectification observed there is very
small.

Finally, we emphasize that long range interactions are not uncommon in physics: they are at the origin of interesting properties and phenomena in classical
and quantum physics, in equilibrium and nonequilibrium phase transitions, etc. They are important at all scales: from cosmology
\cite{vandenA} to nanoscience \cite{French} and quantum information \cite{Koffel, Hauke}. More specifically, in material science, present technology allows to fabricate materials with different properties and even different
types of long range interactions. A noticeable example is given by some Coulomb crystals which are systems intensively investigated and also proposed as candidates for a many-qubit quantum processor. In particular the
engineered Coulomb crystal of ref. \cite{Britton}, given by a trapped beryllium ion system, exhibits Ising like interactions with
$J_{i,j} \propto 1/r_{i,j}^{a}$, where $r_{i,j}$ is the distance between spin pairs and $0<a\leq 3$.
Long-range magnetic dipolar interactions are present in the Ising pyrochlore magnets $Dy_{2}Ti_{2}O_{7}$ and $Ho_{2}Ti_{2}O_{7}$,  which are also of current interest due to the so called spin
ice behavior  \cite{Bra}. An additional example is provided by nanomagnets
such as Permalloy, materials which can be lithographically manipulated \cite{RWang}   in order to display frustrated arrays and exotic states. In such systems
 the interaction typically decays
polynomially as $1/r^{3}$, where $r$ is the distance between the nanodisks.

The system we will discuss here is a one-dimensional chain  of $N$ oscillators described by the Hamiltonian
\begin{equation}\label{Hamiltonian}
H=\sum_{j=1}^{N}\left(\frac{p_j^2}{2m_j}+ \frac{\gamma_{j}q_{j}^{4}}{4}\right) +\sum_{i, j}\frac{\left(q_j - q_{i}\right)^{2}}{2 + 2|i-j|^{\lambda}}~,
\end{equation}
where $q_j$ is the displacement from the equilibrium position of the $j$th particle with mass $m_j$ and momentum $p_j$.
The exponent $\lambda$ gives the decay power of the interparticle interaction, while $\gamma_{j}$ measures the
strength of the on-site potential \cite{Note2}.
The system of anharmonic oscillators is connected to two leads, whose particles have nearest neighbor interactions only so that the heat flux through the leads has the usual simple definition\cite{Zhao98, Lepri, TMai07}.
Two Langevin heat baths \cite{Dharrev}  at different temperatures $T_{L}$ and $T_{H}$ are attached to the two ends of the leads; we take $T_{L}=T(1-\frac{\triangle T}{T})$ and $T_{H}=T(1+\frac{\triangle T}{T})$. (See Fig.1).

The Hamiltonian of the lead reads $H_{lead}=\sum_{i=1}^{n}\left(\frac{p_{i}^{2}}{2m_{i}}+\frac{\gamma_{i}q_{i}^{4}}{4}+\frac{k\left(q_{i+1}-q_{i}\right)^{2}}{2}\right)$,
where $n$ is the number of particles in the lead and $k$ is the nearest
neighbor coupling strength. The main system
is coupled to each lead by a harmonic spring with the same nearest
neighbor coupling strength $k$. In the simulations,
$k=1$ is adopted. We have checked that the main results of this paper
do not depend on the value of $k$ and on the number of
particles in the lead.

\begin{figure}[!]
\includegraphics[scale=0.42]{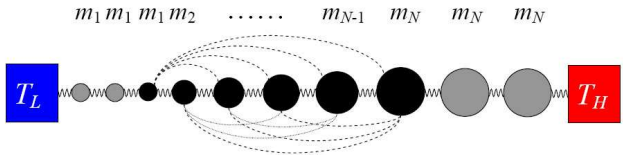}
\vskip-.4cm \caption{(Color online) The schematic plot of our model. At the left and right ends of the chain there are two leads with particles of the same mass $m_1$ (on-site potential strength $\gamma_{1}$) and $m_N$ ($\gamma_{N}$), respectively. Particles in the leads have only nearest neighbor interaction. We have verified that in actual computations it is sufficient to have one particle in each lead.}
\end{figure}

In view of previous results \cite{WPC}, it is expected that in a system with graded mass distribution, e.g.
$m_{1}<m_{2}< \ldots <m_{N}$, thermal rectification will be present, even for the simple case of nearest neighbor interaction
(NN). Now, with long range interactions (LRI), we conjecture that the presence of new
links (interactions) among different sites creates new channels for the heat transport, that favor the heat flow. Besides that, in a graded system,
the new channels connect distant particles with very different masses. Therefore new, asymmetric channels, are created which in turn favors the asymmetric flow, i.e., rectification. Hence, by introducing long range interactions in a graded system, we
expect an increase of the thermal rectification. Moreover, as we increase the system size, new particles
are introduced that, in the case of long range interactions, create new channels for the heat current
 which may avoid the usual decay of rectification with increasing the system length.
  In conclusion, systems with long range interactions may represent a relevant step on the way of constructing efficient thermal diodes.

We would like to remark that the possibility of
rectification growth with long range interactions has been suggested in a recent work \cite{PA2013} where,
 due to analytical difficulties, only an asymptotic study of some simple model has been presented (chains of
oscillators with self-consistent inner reservoirs). Moreover, only infinitesimal differences of temperature are considered,
leading to infinitesimal rectification. Therefore the role of long range interaction in the rectification phenomenon remains
to be understood.

Let us now turn to numerical simulations of our model in order to determine its thermal transport properties  as a function of different parameters such as temperature gradient, mass gradient, etc.  In particular we will compute the rectification factor, defined as $f_{r}=\frac{(J_{+}-J_{-})}{J_{-}}\times100\%$,
where $J_{+}$ and $J_{-}$ represent, respectively, the larger heat
flow and the smaller heat flow, which are measured by inverting the
temperatures of the baths at the two ends of the chain. We found that the larger heat flow $J_{+}$
is obtained when the heat bath at higher temperature is on the side of the heavier mass.

In our simulations the system is evolved for a long enough time to ensure that it has reached
the stationary state. After that the heat current is measured as the average heat flow through the two leads. More specifically,
we verified that the average heat flow into the system is equal to
that out of the system.
We have also performed a series of numerical test in order to ensure that our results do not depend on numerical technical details or on the choice  of the heat baths (for example we have compared with  simulations using Nose-Hoover heat baths and obtained consistent results).
In our simulations, free boundary conditions and velocity-Verlet
algorithm are used. We have compared with simulations using fixed
boundary conditions, and with simulations using Runge-Kutta algorithm
of seventh to eighth order, and obtained consistent results.

As commonly adopted in the literature, we use here the dimensionless units.

We start our numerical studies by investigating the role played by the decay power $\lambda$ of the interparticle interaction. We expect that, as the decay power is increased and, correspondingly,  the interaction
between sites becomes smaller, the rectification factor should decay. Such scenario is confirmed in Fig.2 where we plot the rectification factor as a function on the decay exponent $\lambda$.
Notice however that for $\lambda$ close to one, the rectification factor reaches a maximum value and decreases again when $\lambda$ approaches one. The reason of such behaviour is not clear to us. We surmise that this behaviour is connected to the finite system size and indeed the position of the maximum becomes closer to one as we increase the system size.

\begin{figure}[!]
\includegraphics[scale=0.77]{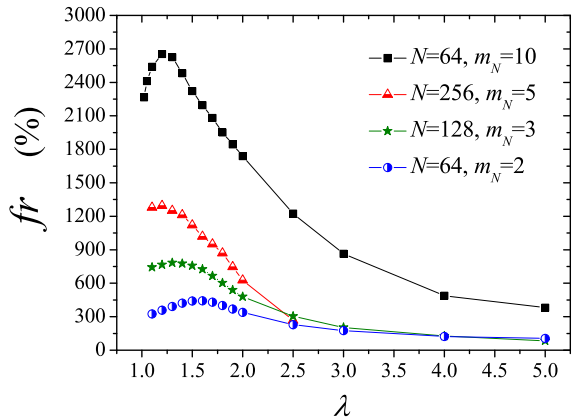}
\vskip-.8cm \caption{(Color online) Rectification factor versus the decay exponent $\lambda$ of the interparticle potential. Here $T=5$, $\frac{\triangle T}{T}=0.9$, $m_{1}=1$ and $\gamma_{j}=1$ for all $j$.
Squares are for $N=64$, $m_{N}=10$; triangles are for $N=256$, $m_{N}=5$; stars are for $N=128$, $m_{N}=3$; and circles are for $N=64$, $m_{N}=2$.}

\end{figure}

We now analyze how the thermal properties depend on the main parameters which determines the asymmetry in the system.
In particular we will compare the behavior of the thermal rectification factor for the LRI-chain and the corresponding chain with nearest
neighbor interaction only. The dependence of $f_{r}$ versus the temperature difference is depicted in Fig.3 for a system with size and mass gradient fixed. The $f_{r}$ amplification due to the presence of LRI is remarkable.

\begin{figure}[!]
\includegraphics[scale=0.77]{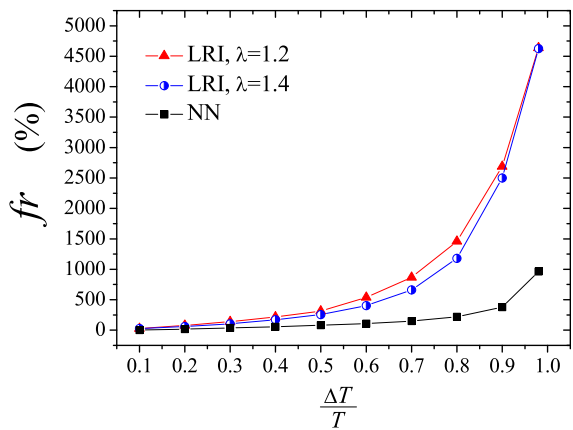}
\vskip-.4cm \caption{(Color online) Dependence of the rectification factor on the temperature gradient.
Here $T=5$, $T_{L}=T(1-\frac{\triangle T}{T})$, $T_{H}=T(1+\frac{\triangle T}{T})$, $N=64$, $m_{1}=1$, $m_{N}=10$. $\gamma_{j}=1$ for all $j$.
Triangles are for LRI with $\lambda = 1.2$, circles are for LRI with $\lambda = 1.4$, squares are for the NN case.}
\end{figure}

As argued before, as we increase the asymmetry in the system, we expect to increase the rectification. The effect is clearly seen  in Fig.4, where  the mass gradient is increased by decreasing $m_1$ with fixed $m_N$.

\begin{figure}[!]
\includegraphics[scale=0.77]{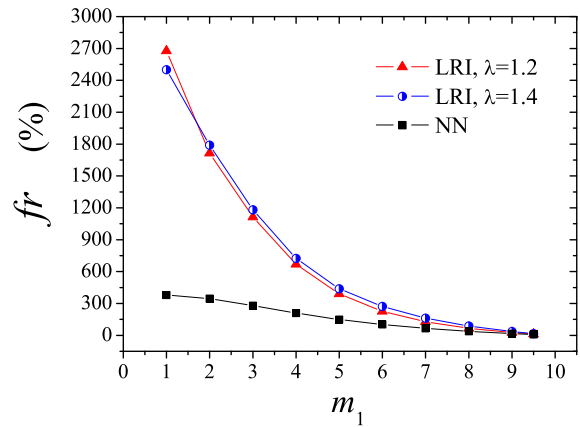}
\vskip-.4cm \caption{(Color online) Dependence of the rectification factor on the mass gradient.
Here $m_{N}=10$, $N=64$, $T=5$, $\frac{\triangle T}{T}=0.9$ and
$\gamma_{j}=1$ for all
$j$. Triangles are for LRI with $\lambda = 1.2$, circles are for LRI with $\lambda = 1.4$, squares are for the NN case.}
\end{figure}

All the above results make transparent the significant gain in the thermal rectification due to long range interactions: for example for the case $T_{H}=9.9$, $T_{L}=0.1$, $N=64$, $m_{1}=1$, $m_{N}=10$, $\lambda = 1.2$, we get the quite large value $f_{r}=4638\%$ \cite{note}.

 We now turn to the investigation of rectification properties as a function of the system size. This is particularly relevant since an unpleasant feature of previous models of rectifiers is the decay of rectification with the increase of the chain length. The results are shown in Fig.5 in which  the mass gradient is fixed.
We
see that the presence of LRI prevents the decay of $f_{r}$ with the system size.
Strictly speaking we cannot make any claim for larger system sizes. However it is clear from Fig.5 that the N-dependence for the LRI case is qualitatively different from the NN case where the decay of the rectification factor with $N$ is immediately evident.

\begin{figure}[!]
\includegraphics[scale=0.77]{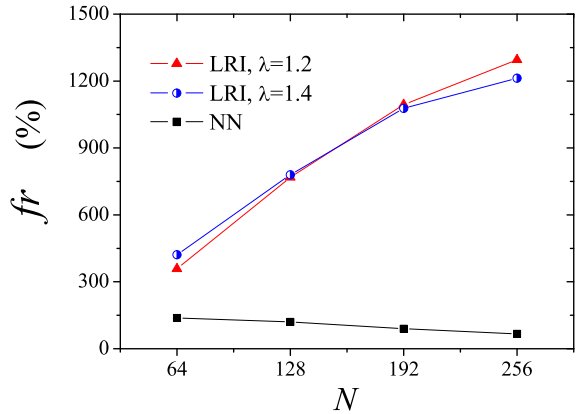}
\vskip-.4cm \caption{(Color online) Dependence of rectification factor on the system size $N$. Here $T=5$, $\frac{\triangle T}{T}=0.9$, $m_{1}=1$ and
$\gamma_{j}=1$ for all $j$. Triangles are for LRI with $\lambda = 1.2$, circles are for LRI with $\lambda = 1.4$, squares are for the NN case. The mass gradient is 1/64. ($m_N=2$ for N=64; $m_N=3$ for N=128; $m_N=4$ for N=192; $m_N =5$ for N=256.).}
\end{figure}

We would like to stress that the rectification effect described here is caused by asymmetry and therefore it should be present in different asymmetric chains.  For example, the rectification phenomenon should be present also in  systems in which the asymmetry in the particle mass distribution is replaced by an asymmetry in the
on-site anharmonic potential.  In order to check this prediction we have considered  our model (1) with equal masses and with
graded on-site potential, namely we took $\gamma_{i}$ linearly increasing with $i$. More precisely $\gamma_{1} = 1 \leq \gamma_{i} \leq \gamma_{max} = \gamma_{N} (i = 1, 2, \ldots, N)$.
The results are shown in Fig.6 which shows a strong rectification effect.
This result is particularly interesting since it shows that rectification may be implemented in a symmetric chain in which
asymmetry is induced by a transverse external field which is graded along the chain.


\begin{figure}[!]
\includegraphics[scale=0.77]{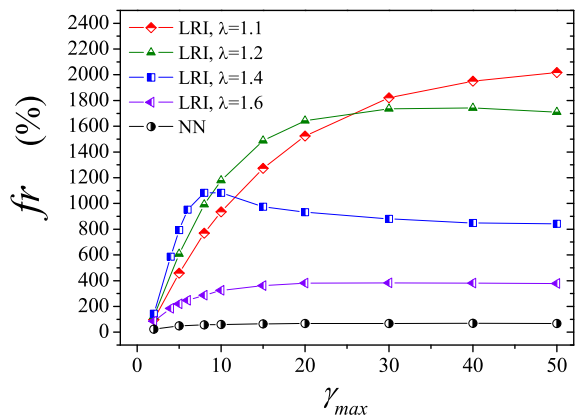}
\vskip-.4cm \caption{(Color online) Rectification factor versus on-site potential coefficient $\gamma_{max}$, for equal-mass system. Size $N=64$, $T=5$, $\frac{\triangle T}{T}=0.9$, $\gamma_{min}=1$. (From particle 1 to particle $N$, $\gamma$ increases linearly. When the hot heat bath is on the $\gamma_{min}$ side, the heat flux is larger.)}
\end{figure}

In this paper we have considered a one dimensional model for reasons of numerical convenience. However the rectification effect should be present in larger dimensional systems  since asymmetry, which is at the root of the rectification mechanisms, can be realized also in two or three dimensions.
In this connection we stress that our results are obtained for graded chains of anharmonic oscillators without any particular requirement or technical restriction.


The main impasse in the development of a more intense experimental research activity on heat current  manipulation is the low rectification factor  which characterizes  current proposals of thermal diodes as well as the fact that the rectification factor decreases with increasing the system size.  In this paper we propose a way to overcome this difficulty.
More precisely,
we show that the presence of long range interactions in a graded mass (or graded structured) system can
considerably increase the rectification factor and avoid its decay as the chain length grows.
 Finally, we stress that the proposed mechanism
is not restricted by geometrical considerations or by space dimensionality.
We are confident that our results will stimulate experimental research on this subject.

 {\bf Acknowledgments:} Useful discussions with Jiao Wang are gratefully acknowledged. E.P. was partially supported by CNPq (Brazil). This work is also supported by MIUR-PRIN.


\begin{thebibliography}{32}
\bibitem{note} We would like to remark that a moderately high rectification, $f_{r} = 350\%$, has been reported \cite{BLi2} in
a quite particular system, namely, a Graphene nanoribbon of a quite specific two-dimensional trapezium shape and in which no
anharmonic on-site potential is present.
\end{thebibliography}
\end{document}